\documentclass[reprint, amsmath,amssymb, showkeys, aps]{revtex4-1}

\usepackage{graphicx}
\usepackage{dcolumn}
\usepackage{bm}

\newcommand{\Comment}[1]{{}}
\usepackage{amsfonts,amsthm,amsmath,amssymb,slashed}
\usepackage[textwidth = 520 pt, textheight = 630 pt]{geometry}

\Comment{\usepackage{color}
\definecolor{MyDarkBlue}{rgb}{0.15,0.15,0.45}
\usepackage[linktocpage=true]{hyperref}
\hypersetup{
colorlinks=true,
citecolor=MyDarkBlue,\bibliography{myref}
linkcolor=MyDarkBlue,
urlcolor=MyDarkBlue,
pdfauthor={Prieslei Goulart},
pdftitle={DRAFT},
pdfsubject={hep-th}
}

\usepackage[numbers,sort&compress]{natbib}
\usepackage{hypernat}}
\usepackage{graphicx}

\newcommand\ignore[1]{}
\def\one{{\,\hbox{1\kern-.8mm l}}}

\newcommand{\Cset}{{\,\,{{{^{_{\pmb{\mid}}}}\kern-.45em{\mathrm C}}}}}

\newcommand{\be}{\begin{equation}}
\newcommand{\bea}{\begin{eqnarray}}

\newcommand{\ee}{\end{equation}}
\newcommand{\eea}{\end{eqnarray}}

\usepackage{color}
\newcommand{\bse}{\begin{subequations}}
\newcommand{\ese}{\end{subequations}}

\begin{document}

\title{Dyonic black holes and dilaton charge in string theory}

\author{Prieslei Goulart}
 \email{prieslei@ift.unesp.br}
 
\affiliation{
Instituto de F\'{i}sica Te\'{o}rica, UNESP-Universidade Estadual Paulista\\
R. Dr. Bento T. Ferraz 271, Bl. II, S\~{a}o Paulo 01140-070, SP, Brazil
}

\date{\today}

\begin{abstract}
We present the four-dimensional non-extremal dyonic black hole solution for Einstein-Maxwell-dilaton theory in absence of a scalar potential written in terms of integration constants only. These integration constants must satisfy a set of conditions imposed by the equations of motion. By defining a posteriori the mass $M$ of the black hole and the dilaton charge $\Sigma$, we show how to recover the dyonic black hole solution found by Kallosh et.al. In particular, our analysis show that there is a possibility in defining whether the dilaton charge or the mass of the black hole is an independent parameter. When the mass of the black hole is the independent parameter, then there is a well-defined limit in which the dilaton charge is zero. For this case, it is straightforward to provide an answer to why is $\phi_{H, \text{extreme}}$ independent of $\phi_{0}$ and to why $\phi_{H, \text{extreme}}=\phi_{0}$ when the dilaton charge is zero.  
\end{abstract}

\keywords{Dyonic black holes; supergravity; dilaton}

\maketitle

\section{Introduction}

Black holes are believed to be the perfect object to address the quantum aspects of gravity. Quantum processes taking place near the horizon, for instance, lead the black hole to evaporate \cite{Hawking:1974sw}. The horizon area surrounding the singularity has a large associated entropy, and this matches the counting of their inner microstates  in the string theory context \cite{Strominger:1996sh}. It is clear then that the knowledge of the full black hole solutions for the low-energy effective actions of string theory (namely supergravity) is crucial for a complete analysis of such phenomena. It turns out that obtaining such solutions is a very difficult task. The reason is that, in general, more realistic supergravity theories contain dilaton fields coupling non-trivially to field strengths, and the equations of motion become highly non-linear. Although some non-extremal solutions were found, much of the progress achieved over the past 25 years was restricted to the treatment of extremal (T=0) black hole solutions. 

This work deals with black hole solutions of the Einstein-Maxwell-dilaton (EMD) theory. The field content of the EMD theory is a metric $g_{\mu\nu}$, a gauge field $A_{\mu}$ and a dilaton $\phi$, and the dilaton couples exponentially to the field strength. Magnetically charged black holes for EMD theory in 4 dimensions were found first in \cite{Gibbons:1984kp, Gibbons:1987ps} and rediscovered later in \cite{Garfinkle:1990qj}. From this solution one obtains the electrically charged one via electric-magnetic duality (S-duality). A more general solution including electric and magnetic charges at the same time, which is referred to as dyonic, was given in \cite{Kallosh:1992ii}. This dyonic solution contains four independent parameters, and it presents a Reissner-Nordstr\"{o}m-like structure in the time and radial components of the metric: it has two horizons and a well-defined zero temperature limit. This is not the case for the magnetic solution, which has a Schwarzschild-like structure in the time and radial components of the metric \cite{Garfinkle:1990qj}.

The inclusion of a scalar field enriches a lot the physics of charged black holes. An intriguing phenomena related to the dilaton field is the attractor mechanism for extremal black holes \cite{Ferrara:1995ih,Strominger:1996kf,Ferrara:1996dd}, which states that the dilaton is attracted to a value on the horizon, $\phi_{H,\text{extreme}}$, that is independent of its value at infinity, namely $\phi_{0}$. The first law of thermodynamics also needed a modification \cite{Gibbons:1996af} to accomodate the dependence upon the moduli $\phi_0$. Also in \cite{Gibbons:1996af}, the authors raised two questions concerning the value of the dilaton on the horizon for extremal black holes: 

(i) Why is $\phi_{H, \text{extreme}}$ independent of $\phi_{0}$?

(ii) Why is $\phi_{H, \text{extreme}}$ given by 
\be \left(\frac{\partial M_{\text{extreme}}}{\partial \phi}\right)_{((p,q), \phi=\phi_{\text{extreme}})}=0? \label{massext}\ee
The second question was answered by the authors in the same refence. It turns out that equation (\ref{massext}) is equivalent to having zero dilaton charge, which implies $\phi_{H, \text{extreme}}=\phi_{0}$. The dyonic solution of reference \cite{Kallosh:1992ii} is not enough to provide an explanation for the first question, which remains unanswered.

In this paper, we provide answers to these two questions. In order to do that, we write the dyonic black hole solutions of EMD theory in terms of integration constants, and then, after writing the solution in terms of the physical charges, we discuss how to recover the dyonic black hole solution found in \cite{Kallosh:1992ii}. By defining the mass or the dilaton charge of the black hole a posteriori, we will see explicitly that there arises a possibility of choosing whether the mass or the dilaton charge is the independent parameter for such solution. This freedom is not present for the extremal solution, and both the dilaton charge and mass depend on the other charges. This dependency of the dilaton charge in terms of the other charges appears explicitly in the extremal solution, and this provides trivial answers to both questions stated above, which are related to the values of the dilaton field on the horizon of the extremal black hole.  

\section{Einstein-Maxwell-dilaton theory}
In order to fix the notation and units we consider the simplest and most general EMD action without a dilaton potential, which can be written as
\be S=\int d^{4}x\sqrt{-g}\left(R-2\partial_{\mu}\phi\partial^{\mu}\phi-W(\phi)F_{\mu\nu}F^{\mu\nu}\right). \label{ad}\ee
Here, $(16\pi G_N)\equiv 1$, where $G_{N}$ is the Newton constant. The field strength is given by
\be F_{\mu\nu}=\partial_{\mu}A_{\nu}-\partial_{\nu}A_{\mu}. \ee
The function $W(\phi)$ will be fixed later. The equations of motion for the metric, dilaton and gauge field, and Bianchi identities, are respectively:
\be R_{\mu\nu}=2\partial_{\mu}\phi\partial_{\nu}\phi-\frac{1}{2}g_{\mu\nu}W(\phi)F_{\rho\sigma}F^{\rho\sigma}+2W(\phi)F_{\mu\rho}
{F_{\nu}}^{\rho}, \label{riccieq}\ee
\be \nabla_{\mu}(\partial^{\mu}\phi)-\frac{1}{4}\frac{\partial W(\phi)}{\partial\phi}F_{\mu\nu}F^{\mu\nu}=0, \label{dileom}\ee
\be \nabla_{\mu}\left(W(\phi)F^{\mu\nu}\right)=0,\label{gaugeeom} \ee
\be \nabla_{\left[\mu\right.}F_{\left.\rho\sigma\right]}=0.\label{bianchi}\ee
In supergravity theories it is common to have more than one scalar field, but all the equations of motion obtained here are easily generalized for these cases. When $W(\phi)=e^{-2\phi}$, (\ref{ad}) is the bosonic sector of $SU(4)$ version of $\mathcal{N}=4$ supergravity theory \cite{Cremmer:1977tt}, for constant axion field. The dyonic black hole solution of \cite{Kallosh:1992ii} was found for the bosonic sector of $SO(4)$ supergravity \cite{Das:1977uy,Cremmer:1977zt,Cremmer:1977tc}, but since the two versions are the same at the level of equations of motion, that is also a solution to the $SU(4)$ version. These are low-energy effective actions describing superstring theory in four dimensions. In this paper we write the dyonic black hole solution to the $SU(4)$ version of $\mathcal{N}=4$ supergravity in terms of integration constants, which differs from \cite{Kallosh:1992ii}, since there the solution is given in terms of physical charges.

\section{General metric and electric-magnetic duality}
The most general form for a static  and spherically symmetric solution is written as
\be ds^{2}=-e^{-\lambda}dt^{2}+e^{\lambda}dr^{2}+C^{2}(r)(d\theta^{2}+\sin^{2}\theta d\phi^{2}),\label{genmet}  \ee
where the metric elements depend on the radial coordinate $r$. Equation (\ref{riccieq}) are non-trivial for the $R_{00}$, $R_{11}$, and $R_{22}$ components (the $R_{33}$ component is the same as the $R_{22}$ component). Adding $R_{00}$ with $R_{11}$
\be -\frac{2C''}{C}=2(\phi')^{2}, \label{segderc} \ee
where the primes denote derivative with respect to the radial coordinate. 
Now, computing $R_{11}$ minus $R_{22}$ we have
\be (e^{-\lambda}C^{2})''=2. \label{segdercl}\ee
We can rewrite $R_{00}$ as 
\be  \frac{d}{dr}\left(C^{2}\frac{d}{dr}(e^{-\lambda})\right)=-2C^{2}e^{-2\phi}(F_{rt}F^{rt}-F_{\theta\phi}F^{\theta\phi}).   \label{niceeq1} \ee
The dilaton equation of motion can be put in the form
\be  \frac{d}{dr}\left(e^{-\lambda}C^{2}\phi'\right)=-C^{2}e^{-2\phi}\left(F_{rt}F^{rt}+
F_{\theta\phi}F^{\theta\phi}\right).  \label{simp2}\ee
By adding and subtracting (\ref{niceeq1}) and (\ref{simp2}) we can form the following equations
\be  \frac{d}{dr}\left(\frac{C^{2}}{2}\frac{d}{dr}(e^{-\lambda})+e^{-\lambda}C^{2}\phi'\right)=-2C^{2}e^{-2\phi}F_{rt}F^{rt}, \label{eqQ}  \ee
\be  \frac{d}{dr}\left(\frac{C^{2}}{2}\frac{d}{dr}(e^{-\lambda})-e^{-\lambda}C^{2}\phi'\right)=2C^{2}e^{-2\phi}
F_{\theta\phi}F^{\theta\phi}. \label{eqP}  \ee
The importance of having the equations written in such a way resides in the fact that we can see explicitly the invariance under S-duality duality transformation,
\be F'^{\mu\nu}=\frac{1}{2\sqrt{-g}}e^{-2\phi}\tilde{\epsilon}^{\mu\nu\rho\sigma}F_{\rho\sigma}, \,\,\, \phi'=-\phi, \label{duality1} \ee
where $\tilde{\epsilon}^{\mu\nu\rho\sigma}$ is the totally antisymmetric Levi-Civita symbol. Just like the Maxwell's equations and Bianchi identities rotate into each other, the same happens to these two equations, showing that the duality transformation is not only an invariance of the gauge field equations but also of the Einstein's and dilaton equations combined. \\

\section{Full dyonic black hole solution}

We are left with a system of equations given by (\ref{gaugeeom}), (\ref{segderc}), (\ref{segdercl}), (\ref{niceeq1}) and (\ref{simp2}). The dyonic black hole solution to the EMD theory is given by
\begin{align}
e^{-\lambda}&=\frac{(r-r_{1})(r-r_{2})}{(r+d_{0})(r+d_{1})}, \,\,\, C^{2}(r)=(r+d_{0})(r+d_{1}), \label{genmets} \\
e^{2\phi}&=e^{2\phi_{0}}\frac{r+d_{1}}{r+d_{0}}, \label{gendil} \\ 
F_{rt}&=\frac{e^{2\phi_{0}}Q}{(r+d_{0})^{2}}, \,\,\, F_{\theta\phi}=P\sin\theta. \label{genele}\end{align}
Here, $Q$ is the electric charge, $P$ is the magnetic charge, $\phi_{0}$ is the value of the dilaton at infinity, and $r_{1}$, $r_{2}$, $d_{0}$, and $d_{1}$ are integration constants which will be fixed. The dilaton equation of motion (\ref{dileom}) implies that all the parameters of the solution must satisfy the following relations
\be (d_{0}-d_{1})[(r_{1}+r_{2})+(d_{0}+d_{1})]=2(e^{2\phi_{0}}Q^{2}-e^{-2\phi_{0}}P^{2}), \label{eq1}\ee
\be (d_{0}-d_{1})(d_{0}d_{1}-r_{1}r_{2})=2(d_{1}e^{2\phi_{0}}Q^{2}-d_{0}e^{-2\phi_{0}}P^{2}),  \label{eq2}\ee
\bea (d_{0}-d_{1})[-(r_{1}+r_{2})d_{0}d_{1}-(d_{0}+d_{1})r_{1}r_{2}]\nonumber \\
=2(d_{1}^{2}e^{2\phi_{0}}Q^{2}-d_{0}^{2}e^{-2\phi_{0}}P^{2}). \label{eq3}\eea
Notice that we can isolate $(r_{1}+r_{2})$ in (\ref{eq1}) and $r_{1}r_{2}$ in (\ref{eq2}) to obtain
\be (r_{1}+r_{2})=2\frac{(e^{2\phi_{0}}Q^{2}-e^{-2\phi_{0}}P^{2})}{(d_{0}-d_{1})}-(d_{0}+d_{1}), \label{sum}\ee
\be r_{1}r_{2}= d_{0}d_{1}-2\frac{(d_{1}e^{2\phi_{0}}Q^{2}-d_{0}e^{-2\phi_{0}}P^{2})}{(d_{0}-d_{1})}.\label{prod}\ee
Replacing (\ref{sum}) and (\ref{prod}) in (\ref{eq3}), we see that this equation is satisfied trivially. In other words, this system of three equations reduces to a system of just two linearly independent equations. Notice that we can combine the equations (\ref{eq1}) and (\ref{eq2} to obtain
\be d_{0}^{2}+(r_{1}+r_{2})d_{0}+r_{1}r_{2}=2e^{2\phi_{0}}Q^{2}, \ee
\be d_{1}^{2}+(r_{1}+r_{2})d_{1}+r_{1}r_{2}=2e^{-2\phi_{0}}P^{2}. \ee
For future convenience, we solve these equations for $d_{0}$ and $d_{1}$ in terms of the other parameters, and obtain
\be d_{0}=\frac{-(r_{1}+r_{2})\pm\sqrt{(r_{1}-r_{2})^{2}+8e^{2\phi_{0}}Q^{2}}}{2}, \label{reld0} \ee
\be d_{1}=\frac{-(r_{1}+r_{2})\pm\sqrt{(r_{1}-r_{2})^{2}+8e^{-2\phi_{0}}P^{2}}}{2}.\label{reld1} \ee
The same procedure can be repeated for equation (\ref{niceeq1}), and the same system of two linearly independente equations can be obtained.

In order to have general expressions, we compute two quantities of interest in terms of the integration constants, which are the temperature of the black hole, given by
\be T=\frac{1}{4\pi}\frac{(r_{2}-r_{1})}{(r_{2}+d_{0})(r_{2}+d_{1})}, \label{temp} \ee
and the dilaton charge, defined as
\be \Sigma=\frac{1}{4\pi}\int d\Sigma^{\mu}\nabla_{\mu}\phi=\frac{(d_{0}-d_{1})}{2}. \label{chargedilaton}\ee
The dilaton charge can be positive or negative, depending on the values of $d_{0}$ and $d_{1}$. The constants $r_{2}$ and $r_{1}$ will be identified with the outer and inner horizon respectively. So far, no boundary condition was imposed on the solution. In the asymptotic region, $g_{tt}$ can be approximated as
\be g_{tt}=-\left(1-\frac{(d_{0}+d_{1}+r_{1}+r_{2})}{r}\right)+\mathcal{O}\left(\frac{1}{r^{2}}\right). \ee
We identify the parameter $M$ as the mass of the black hole, consistent with the weak gravity (Newtonian) limit, i.e. 
\be 2M=(d_{0}+d_{1}+r_{1}+r_{2}). \label{correct}\ee
The Ricci scalar is given by
\be R=\frac{(d_{0}-d_{1})^2(r-r_{1})(r-r_{2})}{2(r+d_{0})^{3}(r+d_{1})^{3}}. \ee
In order to avoid problems with causality, the solution must be restricted to the domain $C^{2}(r)\geq 0$. The singularity is at $r_{S}=-d_{0}$ when $d_{0}> d_{1}$, or at $r_{S}=-d_{1}$ when $d_{1}> d_{0}$.

\section{Relation to known solutions}
In this section we show how to recover the known four parameters non-extremal and extremal black hole solutions found in \cite{Kallosh:1992ii} by Kallosh et al. 
\begin{itemize}
\item Non-extremal dyonic black hole:
\end{itemize}
In order to obtain this solution, we solve (\ref{sum}) and (\ref{prod}) for $r_{1}$ and $r_{2}$, by making the redefinitions $r_{1}\equiv r_{-}$ and $r_{2}\equiv r_{+}$ . The result is
\be r_{\pm}=-\frac{(d_{0}+d_{1})}{2}+\frac{(e^{2\phi_{0}}Q^{2}-e^{-2\phi_{0}}P^{2})}{(d_{0}-d_{1})}\pm\left(\frac{\Delta}{4}\right)^{1/2}, \label{outin} \ee
with
\bea \frac{\Delta}{4}=\left(\frac{d_{0}-d_{1}}{2}\right)^{2}+\frac{(e^{2\phi_{0}}Q^{2}-e^{-2\phi_{0}}P^{2})^{2}}{(d_{0}-d_{1})^{2}}\nonumber \\ -(e^{2\phi_{0}}Q^{2}+e^{-2\phi_{0}}P^{2}). \eea
The definition of mass (\ref{correct}) gives
\be M=\frac{(e^{2\phi_{0}}Q^{2}-e^{-2\phi_{0}}P^{2})}{(d_{0}-d_{1})}. \label{masskall}\ee 
Using also the definition of the dilaton charge, we write this expression as 
\be M\cdot \Sigma=\frac{(e^{\phi_{0}}Q+e^{-\phi_{0}}P)}{\sqrt{2}}\frac{(e^{\phi_{0}}Q-e^{-\phi_{0}}P)}{\sqrt{2}}. \label{ambig}\ee
We will comment on this expression later. Notice that the parameter $\frac{(d_{0}+d_{1})}{2}$ can be removed by shifting the coordinates, i.e.
$r\equiv \rho -\frac{(d_{0}+d_{1})}{2}$. Then the full solution will be written as
\begin{align}
e^{-\lambda}&=\frac{(\rho-\rho_{+})(\rho-\rho_{-})}{(\rho+\Sigma)(\rho-\Sigma)}, \label{metkall}\,\,\, C^{2}(\rho)=(\rho+\Sigma)(\rho-\Sigma),  \\
e^{2\phi}&=e^{2\phi_{0}}\frac{\rho-\Sigma}{\rho+\Sigma}, \label{dilkall} \\ 
F_{\rho t}&=\frac{e^{2\phi_{0}}Q}{(\rho+\Sigma)^{2}}, \,\,\, F_{\theta\phi}=P\sin\theta,\label{gaugekall} \end{align}
with the outer and inner horizons given by
\be \rho_{\pm}=M\pm \sqrt{ \Sigma^{2}+M^{2}-(e^{2\phi_{0}}Q^{2}+e^{-2\phi_{0}}P^{2})}. \ee
This is exactly the solution given in reference \cite{Kallosh:1992ii}. Notice that the dilaton charge given in the same reference is the negative of the one used here. This solution contains four independent parameters: the mass $M$, the electric charge $Q$, the magnetic charge $P$, and the dilaton at infinity $\phi_{0}$. The magnetically charged solutions of \cite{Gibbons:1987ps} and \cite{Garfinkle:1990qj} are obtained from it by setting $Q=0$. By setting further that $P=0$ we recover the Schwarzschild solution.

\begin{itemize}
\item Extremal black holes:
\end{itemize}
When the inner and outer horizon coincide, i.e.  $r_{1}=r_{2}\equiv r_{H}$, we have an extremal ($T=0$) black hole. The horizon is now located at  $r_{H}$, and, from (\ref{reld0}) and (\ref{reld1}) for instance, it is related to the other parameters as
\be r_{H}+d_{0}=\pm\sqrt{2}e^{\phi_{0}}Q,\,\,\, r_{H}+d_{1}=\pm\sqrt{2}e^{-\phi_{0}}P. \ee
The fact that there is more than one possibility of choosing signs will be of relevance for a future analysis. In order to identify this case with the one found in the literature, we make the change of coordinates $\rho = r-r_{H}$
and then the horizon of the extremal black hole is located at $\rho=0$. The extremal black hole solution is then
\begin{align}
ds^{2}&=-e^{-2U}dt^{2}+e^{2U}(d\rho^{2}+\rho^{2}d\Omega^{2}_{2}),  \nonumber \\
e^{2U}&=\left(1\pm\frac{\sqrt{2}e^{\phi_{0}}Q}{\rho}\right)\left(1\pm\frac{\sqrt{2}e^{-\phi_{0}}P}{\rho}\right),  \\
e^{2\phi}&=e^{2\phi_{0}}\frac{(\rho\pm\sqrt{2}e^{-\phi_{0}}P)}{(\rho\pm\sqrt{2}e^{\phi_{0}}Q)}, \label{gendilext} \\
F_{rt}&=\frac{e^{2\phi_{0}}Q}{(\rho\pm\sqrt{2}e^{\phi_{0}}Q)^{2}}, \,\,\, F_{\theta\phi}=P\sin\theta.
\end{align}
We have used isotropic coordinates, i.e. $\rho^{2}=x_{1}^{2}+x_{2}^{2}+x_{3}^{2}$, and consequently $d\rho^{2}+\rho^{2}d\Omega^{2}_{2}=d\vec{x}^{2}$. The parameter $M$ given in (\ref{correct}) can be positive, zero, or negative, depending on the choice of signs and charges in (\ref{reld0}) and (\ref{reld1}). We keep dependency on this arbitrary choice of signs. Using (\ref{chargedilaton}) and (\ref{correct}) we have the following possible configurations 
\be  M=\frac{ e^{\phi_{0}}Q\pm e^{-\phi_{0}}P}{\sqrt{2}}, \,\,\,  \Sigma=\frac{ e^{\phi_{0}}Q\mp e^{-\phi_{0}}P}{\sqrt{2}},\label{signconf1}\ee
\be  M=\frac{e^{-\phi_{0}}P \pm e^{\phi_{0}}Q}{\sqrt{2}}, \,\,\,  \Sigma=\frac{e^{-\phi_{0}}P \mp e^{\phi_{0}}Q}{\sqrt{2}}.\label{signconf2}\ee
When we take the upper sign in (\ref{signconf1}) we recover exactly the extremal dyonic solution found in \cite{Kallosh:1992ii}. But this is just one possibility, since (\ref{signconf1}) and (\ref{signconf2}) show that there are three more. Notice that any configuration of signs respects the product between the mass and the dilaton charge given by equation (\ref{ambig}). It is important to notice that we wrote the explicit dependence of the dilaton field (\ref{gendilext}) in terms of the electric charge $Q$, magnetic charge $P$ and dilaton at infinity $\phi_{0}$. 

\section{Dependent and independent parameters}
In order to recover the solution found by Kallosh et al. \cite{Kallosh:1992ii}, given by (\ref{metkall}), (\ref{dilkall}), and (\ref{gaugekall}), we had to define the mass $M$ using (\ref{masskall}). One could argue that this identification makes it clear that the mass $M$ is a dependent parameter, of the kind $M(Q,P,\phi_{0},\Sigma)$, and the dilaton charge $\Sigma$ is an independent parameter. This argument could be supported by the fact that the horizons (\ref{outin}) blow up in the limit when $d_{0}\rightarrow d_{1}$, i.e. when $\Sigma\rightarrow 0$, resulting in a divergent metric. Moreover, it seems that the zero mass limit is well-defined in (\ref{masskall}). But one could for instance, take the dilaton charge as an independent parameter, and say that (\ref{masskall}) is instead written as
\be \Sigma=\frac{(d_{0}-d_{1})}{2}=\frac{(e^{2\phi_{0}}Q^{2}-e^{-2\phi_{0}}P^{2})}{2M}. \label{sigmakall}\ee
In this picture, the mass now becomes an independent parameter, and the dilaton charge has a dependency on the other parameters of the form $\Sigma(Q,P,\phi_{0},M)$. Moreover, in this picture, the limit when the dilaton charge is zero, i.e. $d_{0}\rightarrow d_{1}$, is well-defined, although the mass $M$ now can not be zero, since this will lead to a divergent metric. This is exactly the case presented by the authors of \cite{Kallosh:1992ii}. In fact, we see clearly that claiming that the dilaton charge is the dependent parameter and the mass is independent is just one possibility, since the equations of motion allow us to choose also the other case. In other words, the only restriction we find is that the product between the mass and the dilaton charge must be given by (\ref{ambig}), which is respected also in the extremal case, as we can see from (\ref{signconf1}) and (\ref{signconf2}). Notice also that the extremal solution of \cite{Kallosh:1992ii} has mass and dilaton charge written as (\ref{signconf1}) with only upper sign. Another possibility would be to take both the mass and dilaton charge as dependent parameters in the non-extremal solution. For this case, one is forced by (\ref{ambig}) to choose one of the possibilities represented by equations (\ref{signconf1}) and (\ref{signconf2}). But for this picture, when both are dependent parameters, one can not escape from the extremal limit, which can be easily seen by inserting $M$ and $\Sigma$ in (\ref{outin}). So, in order to write a non-extremal solution in terms of the physical charges of the black hole, one is forced to choose whether the dilaton charge or the mass is a dependent parameter. 

The advantage of writing the solution in terms of integration constants, instead of physical charges, is beyond the discussion on which parameters are dependent or independent. In the next section, we will see that without defining the integration constants in terms of the mass or dilaton charge, we solve some puzzles related to the dilaton field in the extremal limit, as was stated in the introduction. Moreover, in the picture adopted by Kallosh et al. \cite{Kallosh:1992ii}, we have a description of black holes whose dilaton charge is the dependent parameter, with a well-defined limit when the dilaton charge is zero. In the other picture we take the mass of the black hole as a dependent parameter, with a well-defined zero mass limit. The zero mass limit can not be taken directly from (\ref{sigmakall}), which raises the question of whether this limit really exists or is ill-defined. In fact, it is easy to satisfy equations (\ref{eq1}), (\ref{eq2}), and (\ref{eq3}) at the same time in such a way to construct a massless solution. This massless black hole solution and its physical significance will be discussed in a different paper \cite{Goulart:2016nkv}. Moreover, this solution can be used to construct charged Einstein-Rosen bridges satisfying the null energy condition.

\section{Thermal properties}
The thermodynamical properties follow easily. The general black hole temperature and entropy are given by
\be T=\frac{1}{4\pi}\frac{(r_{+}-r_{-})}{(r_{+}+d_{0})(r_{+}+d_{1})}, \,\,\, S=\pi (r_{+}+d_{0})(r_{+}+d_{1}).  \ee
Notice that the temperature and entropy are written in terms of the integration constants. In the extremal limit, $r_{+}=r_{-}\equiv r_{H}$, the temperature is naturally zero and the entropy is completely independent on the choice of sign configurations in (\ref{signconf1}) and (\ref{signconf2}), i.e.
 \be T=0, \,\,\, S=2\pi QP.  \ee
This also shows that the entropy for extremal black holes is insensitive to boundary conditions i.e. it does not depend on $\phi_{0}$. This is in agreement with the attractor mechanism \cite{Ferrara:1995ih, Ferrara:1996dd}. The entropy was first computed in \cite{Kallosh:1992ii}, and confirmed in several papers dealing also with the attractor mechanism. The value of the dilaton field on the horizon is given by 
\be e^{2\phi_H}=\frac{P}{Q}. \ee
 
The first law of black hole thermodynamics needs a modification to include the dependence upon the moduli $\phi_{0}$. This was done in \cite{Gibbons:1996af} and we write it here. Although we dealt with static solutions, we also include angular momentum for completeness. For fixed mass, $(r_{1}+r_{2})=2M$, this is
\be dM=\frac{\kappa}{8\pi}dA+\Omega dJ+\psi^{\Lambda}dq_{\Lambda}+\chi_{\Lambda}dp^{\Lambda}-\Sigma ^{a}d\phi_{a}, \ee
where $M$ is the mass, $\kappa$ is the surface gravity, $A$ is the area of the horizon, $\Omega$ is the angular velocity, $J$ is the angular momentum, $\psi^{\Lambda}$ and $\chi_{\Lambda}$ are the electrostatic and magnetostatic potentials, $q_{\Lambda}$ and $p^{\Lambda}$ are the electric and magnetic charges, $\Lambda=1,...,n$ is an index labeling all the charges, $\Sigma^{a}$ is the dilaton charge, $\phi_{a}$ is the dilaton at infinity, and $a=1,...,m$ is an index labeling the dilatons evaluated at infinity. In the same reference, it was stated by the authors that two questions arose and motivated the results of their paper. They were rewritten in the introduction of this paper, and as we also mentioned, the authors provided an answer to the second one. It turns out that equation (\ref{massext}) is equivalent to 
\be \Sigma(\phi_{\text{fix}},(p,q))=0. \label{condsigma} \ee
But black holes with vanishing scalar charge must have spatially constant moduli fields: $\phi(r)=\phi_{H, \text{extreme}}=\phi_{0}$. So, for equation (\ref{condsigma}) to be satisfied, we must choose $\phi_{0}$ to be $\phi_{H, \text{extreme}}$. Here, surprisingly, the answers to both questions arise naturally from the solution describing the dilaton field (\ref{gendil}). Because the extremal solution is independent of boundary conditions or any definition of dilaton charge and mass, we can compute $\phi_{H, \text{extreme}}$ from (\ref{gendilext}) without really worrying about how to define mass and dilaton charge. This is achieved just by choosing $\rho=0$, which is the position of the horizon, and this gives
\be e^{2\phi_{H, \text{extreme}}}=\frac{P}{Q}. \label{dilexthor}\ee
We see that the factors of $\phi_{0}$ cancel out of (\ref{gendilext}), which gives an answer to the first question. This miraculous cancellation of $\phi_{0}$ in the solution for the dilaton (\ref{gendilext}) is the root of the attractor mechanism for extremal black holes, and it could only be seen here because we wrote $d_{0}$ and $d_{1}$ in terms of the other parameters. For the second question, we have 
\be \Sigma(\phi_{\text{fix}},(p,q))=0 \Rightarrow d_{0}=d_{1}. \ee
Using the formulas for $d_{0}$ and $d_{1}$, given by (\ref{reld0}) and (\ref{reld1}), this is achieved if the value of the dilaton at infinity is given by
\be e^{2\phi_{0}}=\frac{P}{Q}. \ee
This is the same as (\ref{dilexthor}), which implies $\phi_{H, \text{extreme}}=\phi_{0}$. We see that it is mathematically trivial to answer these questions once the dilaton field is written as (\ref{gendilext}). In other words, we have just written the constants $d_{0}$ and $d_{1}$ appearing in (\ref{gendil}) using (\ref{reld0}) and (\ref{reld1}) in the extremal limit.

\section{Conclusions}
In this paper we studied the dyonic black hole solution of the bosonic part of the $SU(4)$ version of $\mathcal{N}=4$ supergavity with constant axions. We discussed how to obtain the solution found by Kallosh et al. \cite{Kallosh:1992ii}, which is written in terms of the physical charges of the black hole. As the solution presented here is written in terms of the integration constants, we showed that we have a freedom to choose whether the mass or the dilaton charge is the independent parameter. The dilaton field in the extremal case was written with an explicit dependence on the electric and magnetic charges, and the dilaton field at infinity. This allowed us to solve the puzzling questions, raised by the authors of \cite{Gibbons:1996af}, concerning the value of the dilaton on the horizon of the extremal black hole solution.

\begin{acknowledgments}
The author is grateful to George Matsas and Horatiu Nastase for useful discussions. The author is also grateful for the hospitality of the Max-Planck-Institut f\"{u}r  Physik (Werner Heisenberg Institut), where part of this work was developed. This work is supported by FAPESP grant 2013/00140-7 and 2015/17441-5.
\end{acknowledgments}



%

\begin{thebibliography}{14}%
\makeatletter
\providecommand \@ifxundefined [1]{%
 \@ifx{#1\undefined}
}%
\providecommand \@ifnum [1]{%
 \ifnum #1\expandafter \@firstoftwo
 \else \expandafter \@secondoftwo
 \fi
}%
\providecommand \@ifx [1]{%
 \ifx #1\expandafter \@firstoftwo
 \else \expandafter \@secondoftwo
 \fi
}%
\providecommand \natexlab [1]{#1}%
\providecommand \enquote  [1]{``#1''}%
\providecommand \bibnamefont  [1]{#1}%
\providecommand \bibfnamefont [1]{#1}%
\providecommand \citenamefont [1]{#1}%
\providecommand \href@noop [0]{\@secondoftwo}%
\providecommand \href [0]{\begingroup \@sanitize@url \@href}%
\providecommand \@href[1]{\@@startlink{#1}\@@href}%
\providecommand \@@href[1]{\endgroup#1\@@endlink}%
\providecommand \@sanitize@url [0]{\catcode `\\12\catcode `\$12\catcode
  `\&12\catcode `\#12\catcode `\^12\catcode `\_12\catcode `\%12\relax}%
\providecommand \@@startlink[1]{}%
\providecommand \@@endlink[0]{}%
\providecommand \url  [0]{\begingroup\@sanitize@url \@url }%
\providecommand \@url [1]{\endgroup\@href {#1}{\urlprefix }}%
\providecommand \urlprefix  [0]{URL }%
\providecommand \Eprint [0]{\href }%
\providecommand \doibase [0]{http://dx.doi.org/}%
\providecommand \selectlanguage [0]{\@gobble}%
\providecommand \bibinfo  [0]{\@secondoftwo}%
\providecommand \bibfield  [0]{\@secondoftwo}%
\providecommand \translation [1]{[#1]}%
\providecommand \BibitemOpen [0]{}%
\providecommand \bibitemStop [0]{}%
\providecommand \bibitemNoStop [0]{.\EOS\space}%
\providecommand \EOS [0]{\spacefactor3000\relax}%
\providecommand \BibitemShut  [1]{\csname bibitem#1\endcsname}%
\let\auto@bib@innerbib\@empty
\bibitem [{\citenamefont {Hawking}(1975)}]{Hawking:1974sw}%
  \BibitemOpen
  \bibfield  {author} {\bibinfo {author} {\bibfnamefont {S.~W.}\ \bibnamefont
  {Hawking}},\ }\bibfield  {booktitle} {\emph {\bibinfo {booktitle} {{In
  *Gibbons, G.W. (ed.), Hawking, S.W. (ed.): Euclidean quantum gravity*
  167-188}}},\ }\href {\doibase 10.1007/BF02345020} {\bibfield  {journal}
  {\bibinfo  {journal} {Commun. Math. Phys.}\ }\textbf {\bibinfo {volume}
  {43}},\ \bibinfo {pages} {199} (\bibinfo {year} {1975})},\ \bibinfo {note}
  {[,167(1975)]}\BibitemShut {NoStop}%
\bibitem [{\citenamefont {Strominger}\ and\ \citenamefont
  {Vafa}(1996)}]{Strominger:1996sh}%
  \BibitemOpen
  \bibfield  {author} {\bibinfo {author} {\bibfnamefont {A.}~\bibnamefont
  {Strominger}}\ and\ \bibinfo {author} {\bibfnamefont {C.}~\bibnamefont
  {Vafa}},\ }\href {\doibase 10.1016/0370-2693(96)00345-0} {\bibfield
  {journal} {\bibinfo  {journal} {Phys. Lett.}\ }\textbf {\bibinfo {volume}
  {B379}},\ \bibinfo {pages} {99} (\bibinfo {year} {1996})},\ \Eprint
  {http://arxiv.org/abs/hep-th/9601029} {arXiv:hep-th/9601029 [hep-th]}
  \BibitemShut {NoStop}%
\bibitem [{\citenamefont {Gibbons}(1984)}]{Gibbons:1984kp}%
  \BibitemOpen
  \bibfield  {author} {\bibinfo {author} {\bibfnamefont {G.~W.}\ \bibnamefont
  {Gibbons}},\ }in\ \href@noop {} {\emph {\bibinfo {booktitle} {{XV GIFT
  Seminar on Supersymmetry and Supergravity Gerona, Spain, June 4-11, 1984}}}}\
  (\bibinfo {year} {1984})\BibitemShut {NoStop}%
\bibitem [{\citenamefont {Gibbons}\ and\ \citenamefont
  {Maeda}(1988)}]{Gibbons:1987ps}%
  \BibitemOpen
  \bibfield  {author} {\bibinfo {author} {\bibfnamefont {G.~W.}\ \bibnamefont
  {Gibbons}}\ and\ \bibinfo {author} {\bibfnamefont {K.-i.}\ \bibnamefont
  {Maeda}},\ }\href {\doibase 10.1016/0550-3213(88)90006-5} {\bibfield
  {journal} {\bibinfo  {journal} {Nucl. Phys.}\ }\textbf {\bibinfo {volume}
  {B298}},\ \bibinfo {pages} {741} (\bibinfo {year} {1988})}\BibitemShut
  {NoStop}%
\bibitem [{\citenamefont {Garfinkle}\ \emph {et~al.}(1991)\citenamefont
  {Garfinkle}, \citenamefont {Horowitz},\ and\ \citenamefont
  {Strominger}}]{Garfinkle:1990qj}%
  \BibitemOpen
  \bibfield  {author} {\bibinfo {author} {\bibfnamefont {D.}~\bibnamefont
  {Garfinkle}}, \bibinfo {author} {\bibfnamefont {G.~T.}\ \bibnamefont
  {Horowitz}}, \ and\ \bibinfo {author} {\bibfnamefont {A.}~\bibnamefont
  {Strominger}},\ }\href {\doibase 10.1103/PhysRevD.43.3140,
  10.1103/PhysRevD.45.3888} {\bibfield  {journal} {\bibinfo  {journal} {Phys.
  Rev.}\ }\textbf {\bibinfo {volume} {D43}},\ \bibinfo {pages} {3140} (\bibinfo
  {year} {1991})},\ \bibinfo {note} {[Erratum: Phys.
  Rev.D45,3888(1992)]}\BibitemShut {NoStop}%
\bibitem [{\citenamefont {Kallosh}\ \emph {et~al.}(1992)\citenamefont
  {Kallosh}, \citenamefont {Linde}, \citenamefont {Ortin}, \citenamefont
  {Peet},\ and\ \citenamefont {Van~Proeyen}}]{Kallosh:1992ii}%
  \BibitemOpen
  \bibfield  {author} {\bibinfo {author} {\bibfnamefont {R.}~\bibnamefont
  {Kallosh}}, \bibinfo {author} {\bibfnamefont {A.~D.}\ \bibnamefont {Linde}},
  \bibinfo {author} {\bibfnamefont {T.}~\bibnamefont {Ortin}}, \bibinfo
  {author} {\bibfnamefont {A.~W.}\ \bibnamefont {Peet}}, \ and\ \bibinfo
  {author} {\bibfnamefont {A.}~\bibnamefont {Van~Proeyen}},\ }\href {\doibase
  10.1103/PhysRevD.46.5278} {\bibfield  {journal} {\bibinfo  {journal} {Phys.
  Rev.}\ }\textbf {\bibinfo {volume} {D46}},\ \bibinfo {pages} {5278} (\bibinfo
  {year} {1992})},\ \Eprint {http://arxiv.org/abs/hep-th/9205027}
  {arXiv:hep-th/9205027 [hep-th]} \BibitemShut {NoStop}%
\bibitem [{\citenamefont {Ferrara}\ \emph {et~al.}(1995)\citenamefont
  {Ferrara}, \citenamefont {Kallosh},\ and\ \citenamefont
  {Strominger}}]{Ferrara:1995ih}%
  \BibitemOpen
  \bibfield  {author} {\bibinfo {author} {\bibfnamefont {S.}~\bibnamefont
  {Ferrara}}, \bibinfo {author} {\bibfnamefont {R.}~\bibnamefont {Kallosh}}, \
  and\ \bibinfo {author} {\bibfnamefont {A.}~\bibnamefont {Strominger}},\
  }\href {\doibase 10.1103/PhysRevD.52.R5412} {\bibfield  {journal} {\bibinfo
  {journal} {Phys. Rev.}\ }\textbf {\bibinfo {volume} {D52}},\ \bibinfo {pages}
  {5412} (\bibinfo {year} {1995})},\ \Eprint
  {http://arxiv.org/abs/hep-th/9508072} {arXiv:hep-th/9508072 [hep-th]}
  \BibitemShut {NoStop}%
\bibitem [{\citenamefont {Strominger}(1996)}]{Strominger:1996kf}%
  \BibitemOpen
  \bibfield  {author} {\bibinfo {author} {\bibfnamefont {A.}~\bibnamefont
  {Strominger}},\ }\href {\doibase 10.1016/0370-2693(96)00711-3} {\bibfield
  {journal} {\bibinfo  {journal} {Phys. Lett.}\ }\textbf {\bibinfo {volume}
  {B383}},\ \bibinfo {pages} {39} (\bibinfo {year} {1996})},\ \Eprint
  {http://arxiv.org/abs/hep-th/9602111} {arXiv:hep-th/9602111 [hep-th]}
  \BibitemShut {NoStop}%
\bibitem [{\citenamefont {Ferrara}\ and\ \citenamefont
  {Kallosh}(1996)}]{Ferrara:1996dd}%
  \BibitemOpen
  \bibfield  {author} {\bibinfo {author} {\bibfnamefont {S.}~\bibnamefont
  {Ferrara}}\ and\ \bibinfo {author} {\bibfnamefont {R.}~\bibnamefont
  {Kallosh}},\ }\href {\doibase 10.1103/PhysRevD.54.1514} {\bibfield  {journal}
  {\bibinfo  {journal} {Phys. Rev.}\ }\textbf {\bibinfo {volume} {D54}},\
  \bibinfo {pages} {1514} (\bibinfo {year} {1996})},\ \Eprint
  {http://arxiv.org/abs/hep-th/9602136} {arXiv:hep-th/9602136 [hep-th]}
  \BibitemShut {NoStop}%
\bibitem [{\citenamefont {Cremmer}\ \emph {et~al.}(1978)\citenamefont
  {Cremmer}, \citenamefont {Scherk},\ and\ \citenamefont
  {Ferrara}}]{Cremmer:1977tt}%
  \BibitemOpen
  \bibfield  {author} {\bibinfo {author} {\bibfnamefont {E.}~\bibnamefont
  {Cremmer}}, \bibinfo {author} {\bibfnamefont {J.}~\bibnamefont {Scherk}}, \
  and\ \bibinfo {author} {\bibfnamefont {S.}~\bibnamefont {Ferrara}},\ }\href
  {\doibase 10.1016/0370-2693(78)90060-6} {\bibfield  {journal} {\bibinfo
  {journal} {Phys. Lett.}\ }\textbf {\bibinfo {volume} {B74}},\ \bibinfo
  {pages} {61} (\bibinfo {year} {1978})}\BibitemShut {NoStop}%
\bibitem [{\citenamefont {Das}(1977)}]{Das:1977uy}%
  \BibitemOpen
  \bibfield  {author} {\bibinfo {author} {\bibfnamefont {A.~K.}\ \bibnamefont
  {Das}},\ }\href {\doibase 10.1103/PhysRevD.15.2805} {\bibfield  {journal}
  {\bibinfo  {journal} {Phys. Rev.}\ }\textbf {\bibinfo {volume} {D15}},\
  \bibinfo {pages} {2805} (\bibinfo {year} {1977})}\BibitemShut {NoStop}%
\bibitem [{\citenamefont {Cremmer}\ \emph {et~al.}(1977)\citenamefont
  {Cremmer}, \citenamefont {Scherk},\ and\ \citenamefont
  {Ferrara}}]{Cremmer:1977zt}%
  \BibitemOpen
  \bibfield  {author} {\bibinfo {author} {\bibfnamefont {E.}~\bibnamefont
  {Cremmer}}, \bibinfo {author} {\bibfnamefont {J.}~\bibnamefont {Scherk}}, \
  and\ \bibinfo {author} {\bibfnamefont {S.}~\bibnamefont {Ferrara}},\ }\href
  {\doibase 10.1016/0370-2693(77)90277-5} {\bibfield  {journal} {\bibinfo
  {journal} {Phys. Lett.}\ }\textbf {\bibinfo {volume} {B68}},\ \bibinfo
  {pages} {234} (\bibinfo {year} {1977})}\BibitemShut {NoStop}%
\bibitem [{\citenamefont {Cremmer}\ and\ \citenamefont
  {Scherk}(1977)}]{Cremmer:1977tc}%
  \BibitemOpen
  \bibfield  {author} {\bibinfo {author} {\bibfnamefont {E.}~\bibnamefont
  {Cremmer}}\ and\ \bibinfo {author} {\bibfnamefont {J.}~\bibnamefont
  {Scherk}},\ }\href {\doibase 10.1016/0550-3213(77)90214-0} {\bibfield
  {journal} {\bibinfo  {journal} {Nucl. Phys.}\ }\textbf {\bibinfo {volume}
  {B127}},\ \bibinfo {pages} {259} (\bibinfo {year} {1977})}\BibitemShut
  {NoStop}%
\bibitem [{\citenamefont {Goulart}(2016)}]{Goulart:2016nkv}%
  \BibitemOpen
  \bibfield  {author} {\bibinfo {author} {\bibfnamefont {P.}~\bibnamefont
  {Goulart}},\ }\href@noop {} {\  (\bibinfo {year} {2016})},\ \Eprint
  {http://arxiv.org/abs/1611.03164} {arXiv:1611.03164 [hep-th]} \BibitemShut {NoStop}%
\bibitem [{\citenamefont {Gibbons}\ \emph {et~al.}(1996)\citenamefont
  {Gibbons}, \citenamefont {Kallosh},\ and\ \citenamefont
  {Kol}}]{Gibbons:1996af}%
  \BibitemOpen
  \bibfield  {author} {\bibinfo {author} {\bibfnamefont {G.~W.}\ \bibnamefont
  {Gibbons}}, \bibinfo {author} {\bibfnamefont {R.}~\bibnamefont {Kallosh}}, \
  and\ \bibinfo {author} {\bibfnamefont {B.}~\bibnamefont {Kol}},\ }\href
  {\doibase 10.1103/PhysRevLett.77.4992} {\bibfield  {journal} {\bibinfo
  {journal} {Phys. Rev. Lett.}\ }\textbf {\bibinfo {volume} {77}},\ \bibinfo
  {pages} {4992} (\bibinfo {year} {1996})},\ \Eprint
  {http://arxiv.org/abs/hep-th/9607108} {arXiv:hep-th/9607108 [hep-th]}
  \BibitemShut {NoStop}%
\end{thebibliography}
\end{document}